\documentclass[sigconf,authorversion]{acmart}

\AtBeginDocument{%
  }

\usepackage{multirow}
\usepackage{cleveref}
\usepackage{algorithm}
\usepackage{enumitem}
\usepackage{xcolor}        
\usepackage{hyperref}
\usepackage{balance}
\usepackage{comment}
\usepackage{etoolbox}      

\def\equationautorefname~#1\null{Eq.~(#1)\null}

\begin{document}

\title{FIRST: Federated Inference Resource Scheduling Toolkit for Scientific AI Model Access}

\author{Aditya Tanikanti$^1$, Benoit C\^ot\'e$^1$, Yanfei Guo$^1$, Le Chen$^1$, Nickolaus Saint$^2$, Ryan Chard$^1$, Ken Raffenetti$^1$, Rajeev Thakur$^1$, Thomas Uram$^1$, Ian Foster$^{1,2}$, Michael E. Papka$^{1,3}$, Venkatram Vishwanath$^1$}
\affiliation{
$^1$Argonne National Laboratory, Lemont, IL, USA\\
$^2$The University of Chicago, Chicago, IL, USA\\
$^3$University of Illinois Chicago, Chicago, IL
\country{USA}
}

\pagestyle{plain}

\renewcommand{\shortauthors}{Tanikanti et al.}

\begin{abstract}
We present the Federated Inference Resource Scheduling Toolkit (FIRST), a framework enabling Inference-as-a-Service across distributed High-Performance Computing (HPC) clusters. FIRST provides cloud-like access to diverse AI models, like Large Language Models (LLMs), on existing HPC infrastructure. Leveraging Globus Auth and Globus Compute, the system allows researchers to run parallel inference workloads via an OpenAI-compliant API on private, secure environments. This cluster-agnostic API allows requests to be distributed across federated clusters, targeting numerous hosted models. FIRST supports multiple inference backends (e.g., vLLM), auto-scales resources, maintains "hot" nodes for low-latency execution, and offers both high-throughput batch and interactive modes. The framework addresses the growing demand for private, secure, and scalable AI inference in scientific workflows, allowing researchers to generate billions of tokens daily on-premises without relying on commercial cloud infrastructure.
\end{abstract}


\maketitle

\section{Introduction}

AI models are valuable tools in scientific computing, assisting with tasks such as analyzing large datasets \cite{Prince2024} (e.g., genomic sequences, climate patterns), interpreting complex simulations (e.g., particle collisions), generating synthetic data, and optimizing research methodologies. Many scientists at laboratories and universities depend on High-Performance Computing (HPC) clusters, yet deploying scalable, open-source AI inference services on such systems without commercial cloud support is challenging due to complex authentication, scheduling, and resource management. This paper introduces the Federated Inference Resource Scheduling Toolkit (FIRST), which enables Inference-as-a-Service across distributed HPC clusters. It provides cloud-like access to powerful AI models while leveraging existing infrastructure and maintaining the security and scalability requirements of scientific applications.
Our goal is not to compete with cloud providers, but rather to provide an on-premises option to support AI on traditional HPC workloads.

With FIRST, AI models are accessed through an OpenAI-compliant API that is gated with Globus Auth \cite{globus_auth} to accommodate flexible authorization policies. We leverage Globus Compute \cite{10.1145/3369583.3392683} to make our system agnostic to the underlying cluster and scheduler (e.g., Kubernetes, bare metal, PBS, Slurm) and to enable the dispatch of requests to different federated clusters based on customizable criteria such as resource availability. On HPC clusters, compute nodes are acquired dynamically and initialized based on user requests. 
This approach allows FIRST to easily support multiple state-of-the-art inference frameworks (e.g., vLLM \cite{kwon2023efficient}, Infinity \cite{infinity}, SGLang \cite{sglang2024}) and host arbitrary numbers of AI models of any size. 

Once compute nodes are acquired, they are kept ``warm'' to provide a low-latency experience for any user or tool who needs to interact with the hosted models in real time. This is vital for scientific research
and it embeds AI models into existing workflows. For instance, in exploratory data analysis, scientists can use tools with APIs to query genomic datasets on the fly and thus to obtain instant insights into unusual patterns that may guide further investigations. Similarly, during simulation optimization, researchers can interactively adjust parameters, such as those in climate models, and get immediate feedback via integrated platforms. The compute resources can automatically scale up or down to adapt to bursty demands. Our system also provides a batch mode, in which users can submit a large number of inference tasks with a single request, as well as a web interface for interactive chat sessions.

Our work addresses a critical gap in the scientific computing ecosystem, providing an accessible and efficient bridge between traditional HPC infrastructure and the growing demand for AI. 
This approach is an alternative to cloud providers and offers several key advantages. HPC data centers often contain sensitive data that cannot be shared with external providers, and custom models trained on such data can be securely deployed to trusted users. Scientific datasets can be very large (petabytes), and on-premises inference services can naturally interface with such dataset without making any cloud transfer.
Since our framework is designed to be deployed on existing HPC clusters, workloads can scale significantly and become cost-effective for many scientists. Our proof-of-concept deployment on a shared 24-node HPC cluster Sophia~\cite{sophia} processed 8.7 million inference requests, accommodated 76 users, and generated over 10 billion tokens in the past 10 months.

\section{Background and Related Work}
Here we review recent progress in hosting inference services, and discuss the challenges of deploying such services at HPC facilities.

\subsection{Commercial Cloud Solutions}
Major cloud providers now ship inference services that resemble HPC workloads. Azure HPC AI supplies InfiniBand‑connected VMs with AI‑tuned stacks \cite{azure_hpc_ai}; AWS Batch for ML couples SageMaker’s managed serving with elastic batch processing \cite{aws_batch_ml}; Google’s TPU Service exposes high‑bandwidth interconnects and TPU‑pods for low‑latency inference \cite{google_tpu}.  These offerings demonstrate that cloud infrastructures can meet HPC‑scale performance, but they remain external to institutional data‑governance policies.

\subsection{Inference Optimization Frameworks}
A rich ecosystem of libraries accelerates LLM inference on modern accelerators.  NVIDIA’s TensorRT‑LLM \cite{tensorrt_llm} provides kernel‑fusion, quantisation and dynamic‑batching optimisations that can yield roughly 4× the throughput of a vanilla PyTorch implementation.  HuggingFace’s Text Generation Inference (TGI) \cite{hf_tgi} is an open‑source server that supports continuous batching and efficient attention mechanisms.  The lightweight C++ engine llama.cpp \cite{dettmers2022llm} enables 8‑bit quantised LLaMA models to run on commodity hardware.  vLLM \cite{kwon2023efficient} introduces PagedAttention together with continuous batching to maximise GPU memory utilisation.  SGLang \cite{sglang2024}, a domain‑specific language for LLM serving, adds RadixAttention and other specialised optimisations, achieving up to 3.1× the throughput of vLLM on selected models.  Finally, Infinity \cite{infinity} builds on FlashAttention‑2 \cite{zhang2023flashattention} and aggressive memory‑management strategies to deliver high‑throughput, low‑latency serving.  Together these libraries constitute the core building blocks for modern inference services, although they generally assume a traditional, always‑on server deployment model. 

\subsection{Inference Services at HPC Facilities}
\label{sec:inference_at_hpc}
Integrating the above kernels into HPC centres is non‑trivial.  Production HPC systems rely on batch schedulers (Slurm \cite{slurm}, PBS \cite{pbs_pro}, LSF \cite{lsf}) that optimise resource allocation for long‑running jobs, yet they are ill‑suited for the low‑latency, interactive patterns typical of inference \cite{10.1145/3581784.3607067}.  Consequently, most existing solutions require users to manually provision and configure servers, creating a steep barrier for domain scientists.

Project Garden \cite{gardens_ai} supplies code for cross‑DOI AI access but lacks OpenAI‑compatible APIs and does not exploit HPC‑scale resource management.  Commercial batch services (AWS, Azure) \cite{aws_batch_ml,azure_hpc_ai} offer elasticity but operate outside the secure on‑premises environment required by many scientific workflows.  Open‑source “llm‑serving’’ frameworks \cite{llm-serving} expose APIs on dedicated nodes yet provide no dynamic model registration, scheduler integration, nor institutional authentication.

Our contribution fills this gap: a service that couples cloud‑style APIs with native HPC schedulers, offering both interactive and batch inference while respecting the security and policy constraints of scientific facilities.

\subsection{Scientific Computing with Globus}
Globus \cite{5755602} has established itself as a critical platform for scientific workflows, providing authentication, transfer, and sharing capabilities across institutional boundaries. Globus Compute~\cite{10.1145/3369583.3392683}, in particular, enables Function-as-a-Service capabilities across distributed computing resources and can seamlessly interface with HPC clusters and job schedulers. These advances provide a foundation for building federated services that leverage authentication tools such as Globus Auth~\cite{globus_auth} to conform with institutional policies.

\section{FIRST Architecture}
\label{sec:first_architecture}

FIRST consists of three main components: 1) an Inference Gateway API to process user requests, 2) Globus Compute to execute tasks on HPC resources, and 3) Model Serving Tools to efficiently perform the LLM inference. This layered architecture separates concerns and enables flexible deployment across different types of environments. 
The following subsections describe the role of each component.


\subsection{Inference Gateway API}

The Inference Gateway consists of an asynchronous Django-Ninja \cite{django_ninja_website} API application that connects to a PostgreSQL database and communicates with Globus.
The Gateway is deployed in a virtual machine environment, separate from the HPC clusters where the LLMs are hosted. The API application is served by Gunicorn~\cite{gunicorn} via Uvicorn~\cite{uvicorn} workers, and can be deployed with \texttt{systemctl} services or with containers. An Nginx~\cite{nginx_inc} web server sits between users and the application, which encrypts traffic with SSL certificates. 

\subsubsection{\textbf{Role and Responsibilities}}
The Gateway is the main entry point for users to interact with the system. The API is OpenAI-compatible and supports the chat completions, completions, embeddings endpoints. It is also responsible for validating user identities, validating incoming data, protecting services by rate limiting and caching, converting user requests into Globus Compute requests (see \autoref{sec:globus_compute}), and logging all user activities in the PostgreSQL database. The metrics layer provides real-time monitoring of the compute resources and queue status. Performance and summary metrics are also exposed through a web dashboard.

\subsubsection{\textbf{Authentication and Authorization}}
The Gateway API authorization layer was developed as a resource server on Globus Auth, an OAuth2 and OpenID compliant identity and access management service. Globus Auth allows users to login from different institutions across the world with multi-factor authentication. The API uses Globus policies to control access to the platform and secure the HPC resources. It also uses Globus Groups to implement role-based access control for maintaining security in multi-user HPC environments that may have varying levels of access. For example, researchers working on sensitive projects may be granted special access to specific models or computational resources.


The Gateway API secures requests via Globus Auth access tokens obtained through an authorized identity provider. 
These tokens are passed in the request headers and cached for rapid repeated requests. This architecture allows the Gateway API to efficiently utilize powerful API models at high request cadence, while maintaining a secure environment.



\subsection{Globus Compute}
\label{sec:globus_compute}
Globus Compute \cite{10.1145/3369583.3392683} is a Function-as-a-Service framework that can remotely trigger compute tasks on any system where an \textit{endpoint} is deployed. In FIRST, it acts as the communication layer between the Gateway API and the HPC resources. This service is hosted at AWS, is fully maintained and supported by the Globus team, and naturally integrates with Globus Auth to enhance security.

\subsubsection{\textbf{Role and Responsibilities}}
At the Gateway API level, the Globus Compute SDK forwards each inference request to the Globus Compute service in the cloud. The request carries (i) the pre‑registered inference function to be invoked and (ii) the identifier of the Compute endpoint on which the function should run. Upon receipt, the Globus service validates the request against the access‑policy defined in \autoref{sec:gc_access_permissions}, dispatches the function to the chosen endpoint, and finally relays the inference result back to the Gateway host for delivery to the user.

Within each HPC facility, FIRST administrators deploy a Globus Compute endpoint on the target cluster(s). Each endpoint is configured independently (\autoref{sec:gc_endpoint_config}) to host one or more LLMs, with the specific models selected according to their size and the available compute nodes. When a request arrives, the endpoint automatically acquires the necessary resources—either on local nodes, inside a Kubernetes pod, or through a batch‑scheduler submission (e.g., PBS or Slurm)—executes the inference, and returns the output to the Globus service without human intervention.

Because the Gateway API is agnostic to the underlying execution environment, administrators retain full control over where and how inference is performed. This flexibility enables FIRST to support an arbitrary number of LLMs, employ diverse inference frameworks, and distribute load across multiple federated HPC clusters, all without requiring any changes to the Gateway API code.

\subsubsection{\textbf{Endpoint Configuration and Safety Features}}
\label{sec:gc_endpoint_config}
The flexible Globus Compute endpoint configuration can adapt to different hardware environments, and is designed to ensure efficient resource utilization, fault tolerance, and optimal performance. Key features of such a configuration include:

\textbf{Auto-scaling}: The system automatically provisions additional compute resources when demand exceeds capacity. Administrators can set the maximum number of parallel inference tasks within a node, and the maximum number of nodes an LLM can scale up to.

\textbf{Fault Tolerance}: 
Process management scripts monitor inference server health and automatically restart failed instances. 

\begin{figure}[h]
    \centering
    \includegraphics[width=\columnwidth]{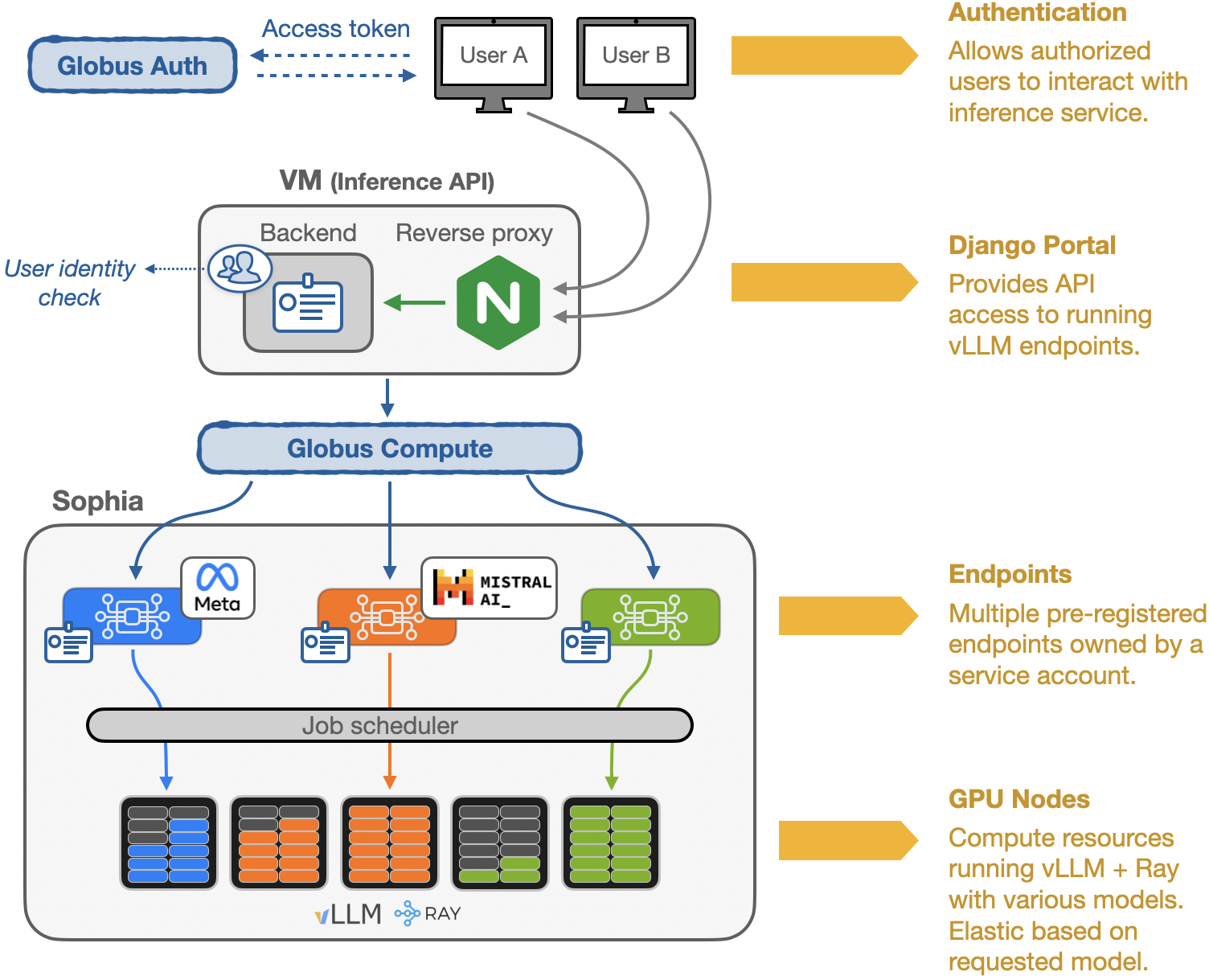}
      \vspace{-0.1in}
    \caption{High-level overview of the FIRST framework integration with ALCF infrastructure and Globus services.}
    \label{fig:alcf_framework}
\end{figure}

\textbf{Resource Utilization}: 
Multiple models can be co-located on a single node to promote efficient use of GPU resources. For example, in a compute node with 8 GPUs, a 70B parameter model might use 6 GPUs, while 8B and 7B models use the remaining 2 GPUs. 

\textbf{Hot Node Management}: 
Nodes remain active after completing requests, allowing subsequent requests 
to be processed without model reloading overhead. This strategy significantly reduces latency for interactive workloads while still allowing resources to be released after an extended idle period (currently 2 hours).

\textbf{Security}: 
Only functions that are pre-registered by the administrators are permitted to be executed on an endpoint, preventing execution of malicious code.

\subsubsection{\textbf{Access and Permissions}}
\label{sec:gc_access_permissions}
The Globus Compute SDK and endpoints can be authenticated with a common Globus confidential client, which is owned by the system administrators. With the client identification and secrets securely stored and hidden, general users cannot communicate with the Globus Compute endpoints directly. Instead, they must go via the Inference Gateway API and all validation steps in order to interface with the running LLMs. 

\subsection{Model Serving Tools}
The Serving Tools represent the LLM serving frameworks that Globus Compute deploys upon initialization to serve models (see \autoref{fig:alcf_framework}). 
Our service is currently configured for serving open-source LLMs, and currently supports vLLM and Infinity, which are optimized for NVIDIA A100, H100, and AMD MI250 GPUs. However, other optimization frameworks can be deployed and configured according to the HPC cluster hardware.

\section{Deployment of the Service}
We describe how we deployed the inference service on an HPC system. \autoref{fig:alcf_framework} shows the overall architecture, with the Gateway API hosted on a virtual machine, and the LLMs served through Globus Compute endpoints deployed on the Sophia cluster~\cite{alcf_sophia_docs} at the Argonne Leadership Computing Facility (ALCF).


FIRST provides a comprehensive set of OpenAI-compliant API endpoints.
This compatibility ensures that researchers can leverage their existing codebase and tools without significant adaptation efforts. The system also supports multiple inference frameworks (see \autoref{sec:model_serving_framework_selection}) and a wide range of models, from small language models to large vision-language models (see \autoref{sec:available_models}). We describe the user interaction in \autoref{sec:user_interaction}, discuss wait times for interactive requests in \autoref{sec:performance_wait_times}, and highlight the batch mode in \autoref{sec:batch_processing}.
Since its deployment on July 2024, this service has processed 8.7 million inference tasks, served 76 users in total, and generated over 10 billion tokens. About 4.1~million tasks were single interactive requests. The other 4.6~million tasks
were packaged into 49 batch requests, which generated over 6.2 billion tokens. 



\subsection{Model Serving Framework Selection}
\label{sec:model_serving_framework_selection}
Numerous benchmark studies, such as LLM-Inference-Bench\cite{10.1109/SCW63240.2024.00178}, compare the performance of various inference serving backends across different hardware architectures. For the initial deployment of FIRST, we selected vLLM \cite{kwon2023efficient} as the primary inference framework. This decision was based on findings available at the time, which indicated vLLM offered superior throughput and low latency, largely attributed to its PagedAttention mechanism \cite{runpod_vllm_blog}. Furthermore, vLLM's compatibility with a range of hardware, including GPUs from NVIDIA, AMD, and potentially others, aligned with our goal of a flexible HPC service. 

While frameworks such as NVIDIA's TensorRT-LLM \cite{tensorrt_llm} may offer higher performance on specific NVIDIA hardware, a key design principle of FIRST is its backend agnosticism. Our architecture can readily integrate with any of the inference frameworks discussed in Section 2.2 (e.g., TensorRT-LLM, TGI \cite{hf_tgi}, SGLang \cite{sglang2024}), provided they expose an OpenAI-compatible API. 


\subsection{Available Models}
\label{sec:available_models}
FIRST exposes a curated set of models covering three functional groups.  For chat‑language tasks we provide several families: Qwen2.5 (14B,7B,32B), Meta‑Llama 3/3.1 (70B,405B), Mistral‑7B‑Instruct‑v0.3 and Mixtral‑8x22B‑Instruct‑v0.1, and the science‑focused AuroraGPT suite \cite{auroragpt_website} (7B,IT‑v4‑0125,Tulu3‑SFT‑0125).

Multimodal vision‑language models are also available, namely Qwen/Qwen2‑VL‑72B‑Instruct and Meta‑Llama‑3.2‑90B‑Vision, enabling combined text‑image interactions.  For embedding generation we ship NVIDIA’s NV‑Embed‑v2, which yields high‑quality vector representations for downstream retrieval‑augmented pipelines.  Adding a new model is straightforward: the model only needs to be supported by one of the configured back‑ends (e.g., vLLM, Infinity), after which it can be registered via the service’s dashboard.  The current catalogue reflects a mix of user‑driven demand and usage‑pattern insights gathered from our monitoring infrastructure.

\subsection{Performance and Wait Times}
\label{sec:performance_wait_times}

FIRST balances responsiveness with HPC resource efficiency. When a user requests a model that is already loaded onto GPUs (a ``hot'' model), inference requests are processed with minimal latency, enabling real-time interaction. However, if the requested model is not currently active (a ``cold'' start), FIRST initiates a process that involves several steps typical of HPC environments. First, a job is submitted to the cluster's scheduler (e.g., PBS). This job may spend time waiting in the queue until the necessary compute nodes and GPUs are allocated. Once allocated, the primary task is loading the model's weights from storage into the GPU memory. The time required for this loading process and the resulting memory footprint are highly dependent on the model's parameter count as well as on parameters such as the chosen context size, applied quantization techniques (if any), and expected sequence lengths. For instance, an 8B parameter model (typically requiring around 16~GB of VRAM) generally has a smaller memory footprint and loads relatively quickly onto fewer GPUs compared to a large 405B parameter model (potentially needing 800~GB or more of VRAM), which requires coordinating the loading process across multiple nodes and GPUs, significantly increasing the cold start time.

To provide users transparency into model availability and potential wait times, FIRST offers a \textbf{`/jobs'} API endpoint. This endpoint queries the underlying scheduler and reports the status of models managed by the service. Users can see which models are currently ``running'' (hot and ready for immediate inference), ``starting'' (nodes acquired, model loading in progress), or ``queued'' (waiting for resource allocation). 


\subsection{Batch Processing}
\label{sec:batch_processing}

For large-scale inference tasks, such as for generating comprehensive textual descriptions of genomic features
, FIRST provides a dedicated batch processing mode. This mode achieves higher throughput compared to interactive requests by eliminating the overhead associated with online model serving. Online serving often involves requests passing through a shared server process, which can face limitations (e.g., vLLM's API server historically being single-threaded \cite{vllm_issue_12705}) and resource contention with other users. 

In contrast, FIRST's batch mode executes each batch job as a dedicated HPC job. This job loads the specified model solely for that task, processing all requests from the user's input file directly without the mediation of a shared online server. Users submit batch jobs via the `/v1/batches' endpoint, providing an input file in JSON Lines format where each line constitutes a complete inference request. This approach allows for the efficient processing of thousands of requests within a single, dedicated job allocation, maximizing resource utilization for the task. The batch system 
also provides status updates for monitoring long-running jobs.

\subsection{Federation Implementation}
The federation layer is a key advantage of our architecture, enabling seamless access to resources across distributed clusters. As a proof of concept, we deployed additional endpoints on the ALCF Polaris System \cite{alcf_polaris_docs} - another HPC system at ALCF. We then implemented a development API URL that does not target any specific cluster and is agnostic. Once a request is received, our API queries the database to see which clusters can host the inference (Sophia or Polaris in our case), queries the publicly available status of each cluster, decides which cluster to use based on node availability, and submits the request to the automatically selected cluster. The configuration of this functionality needs to be adapted for deployments at other facilities, since it relies on how the status of compute resources is queried internally within each facility. 

The core logic for selecting an endpoint in this federated proof-of-concept follows a simplified priority-based algorithm. It first attempts to find an endpoint where the requested model is already running or queued. If none exists, it searches for an endpoint associated with a cluster that has available nodes. If neither condition is met, it defaults to the first endpoint configured for that model, where priority is determined simply by the order in which endpoints are listed in the configuration registry.
This ensures that requests are preferentially sent to active instances for low latency.
\subsection{User Interaction}
\label{sec:user_interaction}
Users interact with the system through a streamlined authentication and API workflow designed to abstract complexity while maintaining security (see \autoref{fig:alcf_framework}). The process begins with authentication through Globus Auth, which integrates with institutional identity providers, ensuring that only authorized users can access the system. Access tokens are valid for 48 hours and can be automatically refreshed to reduce the need for frequent re-authentications. Once authenticated, users can make API requests using standard HTTP clients or the OpenAI Python package. This allows researchers to leverage familiar tools and libraries, reducing the learning curve associated with adopting new systems. We provide an easy-to-use Python script to guide users through the authentication, as well as numerous examples on how to interact with our system.

\subsection{Web Interface for Interactive Chat}
\label{sec:user_chat_interface}
\begin{figure}[h]
    \centering
    {\setlength{\fboxsep}{0pt}\fbox{\includegraphics[width=0.4\textwidth]{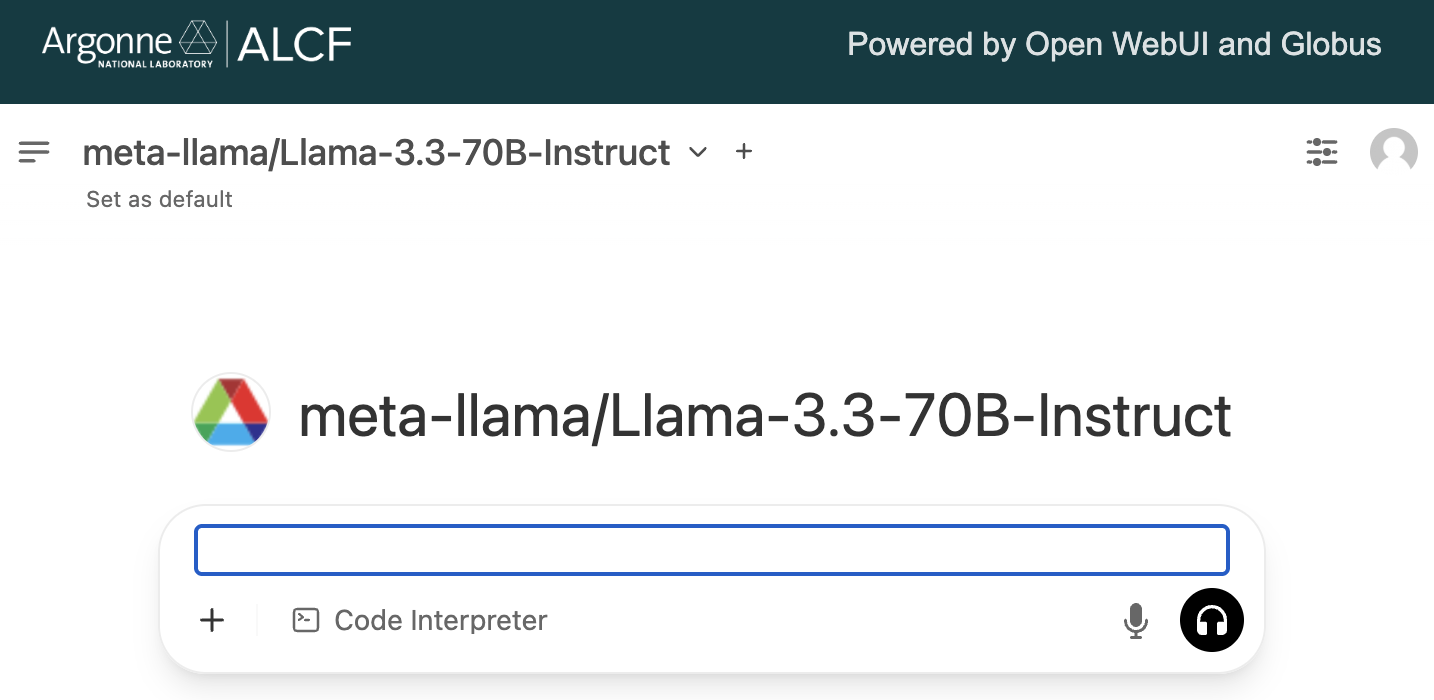}}}
      \vspace{-0.1in}
    \caption{Inference service user chat interface.}
    \label{fig:webui}
\end{figure}

We have developed a web interface that enables users to engage in interactive chat sessions with models running on the backend (see \autoref{fig:webui}). The interface is based on Open WebUI~\cite{openwebui}, customized to integrate with Globus Auth for secure access, and is supported by a backend stack consisting of FastAPI served by Uvicorn workers behind an Nginx reverse proxy, with a PostgreSQL database used to persist user sessions and request metadata. All user requests, along with the access tokens generated through the authentication process, are forwarded to our Gateway API. Through the interface, users can select and view which models are currently running from a dropdown menu, maintain chat histories, compare responses from different LLMs in a multi-column layout, and adjust OpenAI-compliant parameters. The interface also supports streaming responses and rich formatting for equations, code, and tables, making the user experience both functional and visually engaging.

\section{Performance Evaluation}
%

We conducted benchmarks to highlight the concurrency levels achievable with the FIRST framework. Our goal here is not to compete with commercial cloud providers, but rather to measure its performance when deployed on limited HPC resources.

\begin{figure*}[hbt!]
  \centering
  \includegraphics[width=0.8\textwidth]{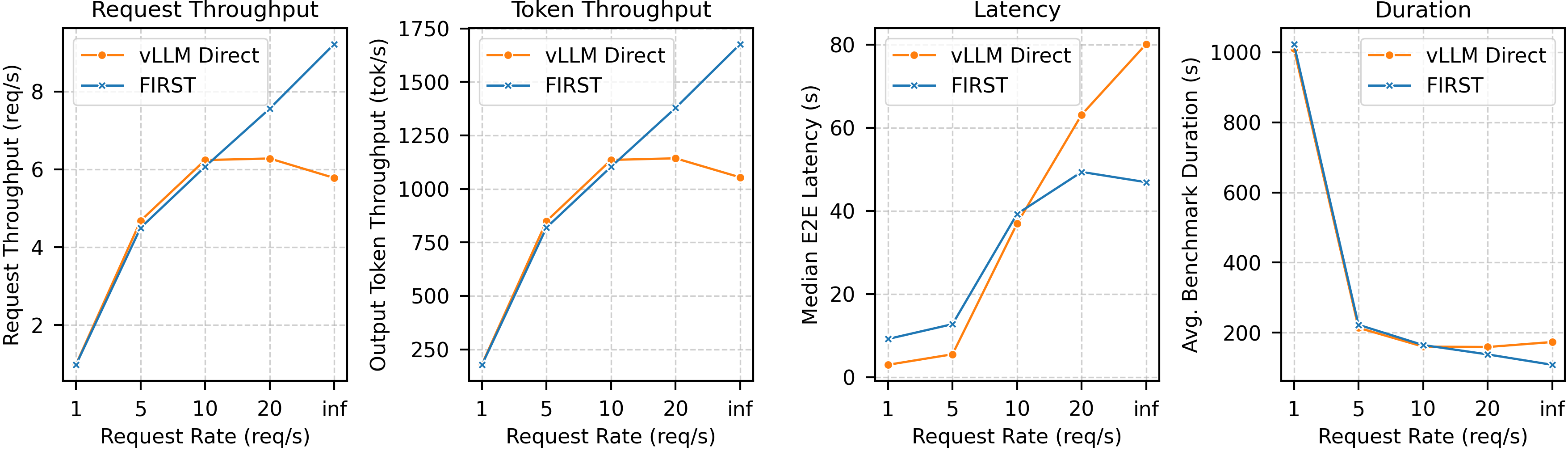}
    \vspace{-0.1in}
  \caption{Performance vs.\ request rate for FIRST vs.\ vLLM Direct for Llama 3.3 70B on a single Sophia node with 8 A100 GPUs, showing request throughput, output token throughput, median end-to-end latency, and duration vs.\ offered request rate.}
  \label{fig:rate_comparison_70b}
\end{figure*}

\subsection{Performance Metrics}
We focus on the following key metrics \cite{Sagi2025} to evaluate performance:
\begin{itemize}[leftmargin=*]
    \item \textbf{Request throughput (req/s):} The number of successful inference requests completed per second. This metric measures the overall request processing capacity of the system.
    \item \textbf{Output Token Throughput (tok/s):} The total number of output tokens generated by the model per second across all successful requests. This metric reflects the effective generation speed.
    \item \textbf{Median End-to-End Latency (s):} The time elapsed from sending a request to receiving the complete response. We primarily report the median latency, as it is less sensitive to outliers than the mean. This metric reflects the user-perceived responsiveness for a single request.


    \item \textbf{Benchmark Duration (s):} The total time taken to complete all requests in a given benchmark run. This provides an overall measure of how quickly the entire workload was processed.
\end{itemize}


\subsection{Experimental Setup}
\subsubsection{\textbf{Hardware}}
All benchmarks were conducted on Sophia at ALCF \cite{alcf_sophia_docs}. Sophia comprises 24 NVIDIA DGX A100 nodes, each equipped with eight NVIDIA A100 Tensor Core GPUs (mostly 40~GB, with two nodes offering 80~GB GPUs for a total of 8320~GB VRAM across the system) and two AMD Rome CPUs. The nodes feature fast local storage (15~TB SSD per node) and are interconnected with a Mellanox HDR Infiniband fabric in a fat-tree topology. Specific model configurations using Tensor Parallelism (TP) were 4 GPUs, TP=4 for \textbf{Llama 3.1 8B} and 8 GPUs, TP=8
for \textbf{Llama 3.3 70B}.

\subsubsection{\textbf{Software}} The following software setup was used.
\begin{itemize}[leftmargin=*]
    \item \textbf{Inference Server:} vLLM (version 0.8.2) was used for the inference engine for both Direct and FIRST benchmarks.
    \item \textbf{FIRST Deployment:} The FIRST API Gateway was deployed using Gunicorn with Uvicorn workers, configured to handle asynchronous tasks. The specific Gunicorn configuration utilized \texttt{multiprocessing.cpu\_count() $\times$ 2 + 1} workers, 4 threads per worker, and a backlog of 2048.
    \item \textbf{Benchmarking Tool:} We adapted the benchmark script provided by the vLLM project \cite{vllm_benchmark_script_github} to measure performance. This script utilizes the ShareGPT dataset \cite{sharegpt2023}, which contains thousands of real-world user-AI conversations across diverse topics. Using ShareGPT ensures our benchmarks reflect realistic usage patterns and query types, providing a robust evaluation of throughput and latency metrics.
    \item \textbf{Workload:} The benchmarks used the ShareGPT dataset, sending 1000 requests. For peak performance tests, an ``infinite'' request rate was used, meaning all requests are sent as quickly as possible at the beginning of the benchmark to fully saturate the server. Using an infinite request rate effectively emulates the usage patterns of popular open models where numerous users might send requests concurrently. To ensure fair comparison between scenarios (FIRST vs. Direct, different request rates), the same set of input prompts and corresponding target output lengths were used for each model across all relevant tests.
    \item \textbf{OpenAI Models:} We ran benchmarks against the OpenAI API for GPT-4o-mini for comparison, using the same benchmarking tool and workload where applicable (subject to API rate limits).
\end{itemize}
\subsubsection{\textbf{Benchmark Scenarios}} For all scenarios involving vLLM (Direct and FIRST), identical vLLM serving parameters were used for a given model to ensure consistency.
\begin{itemize}[leftmargin=*]
    \item \textbf{vLLM Direct:} Requests were sent directly from the benchmarking client to the vLLM OpenAI-compatible API endpoint running on the designated Sophia nodes.
    \item \textbf{FIRST:} Requests were sent to the FIRST API Gateway, which then authenticated the request, routed it to the appropriate Globus Compute endpoint on Sophia, executed the inference function, and returned the result.
    \item \textbf{OpenAI API:} Requests were sent to the official OpenAI API endpoints for GPT-4o-mini.
\end{itemize}

\subsection{Results and Discussion}

\subsubsection{\textbf{Performance vs. Request Rate (Llama 3.3 70B Model)}}
To understand how FIRST behaves under varying load, we benchmarked the Llama 3.3 70B model at different request rates (1, 5, 10, 20 req/s, and infinite/maximum rate) and this is depicted in ~\Cref{fig:rate_comparison_70b}.


At lower request rates (1 and 5 req/s), the additional network hop and processing within FIRST introduces observable latency compared to direct vLLM access (e.g., median latency of 9.2s vs.\ 3.0s at 1 req/s). However, as the request rate increases (10 req/s and beyond), FIRST demonstrates superior scalability. At 20 req/s and the infinite rate, FIRST achieves significantly higher request and output token throughput (e.g., 9.2 vs.\ 5.8 req/s and 1677 vs.\ 1054 tok/s). On a related note, FIRST's median latency at these high rates becomes lower than direct access (46.9s vs.\ 80.2s at infinite rate).

This scaling advantage stems from FIRST's architecture. The asynchronous API gateway (using Gunicorn/Uvicorn) effectively buffers and manages high volumes of incoming user requests, preventing the backend inference engine from being overwhelmed. This contrasts with directly exposing the vLLM server, which historically faced limitations in its own API handling capacity (e.g., single-threaded processing \cite{vllm_issue_12705}). This demonstrates that our gateway architecture can provide superior performance under high load conditions, validating our approach for production HPC environments. FIRST provides this scaling capability transparently to users, without requiring changes to their client-side scripts. We implemented three sets of improvements to reach this level of efficiency:

\textbf{Optimization 1}: We initially determined request status by polling every 2~s, returning results to users once tasks were labeled as completed. We optimized this by using concurrent Python \texttt{future} objects to retrieve results as soon as possible, which eliminated the polling overhead for each request.

\textbf{Optimization 2}: In our early design, each request queried the Globus servers to introspect the user token and to create new connections with the compute endpoints.  To improve efficiency, these repetitive steps are now cached for frequently incoming requests. This eliminated 2~s from the latency of each request and prevented our framework from being rate-limited by the Globus services.

\textbf{Optimization 3}: A major improvement was the migration from synchronous Django REST~\cite{django-rest-framework} to asynchronous Django Ninja~\cite{django_ninja_website}. Originally, only nine requests could be processed at a time, and the API's compute resources were idle while waiting for results. Now, the API can continuously offload requests to the HPC cluster. Benchmarks with Artillery~\cite{artillery} (100 incoming requests per second for 300 seconds) showed that over 8000 inference tasks could be queued at Globus, successfully removing our API as a bottleneck. The benchmark also showed that the response throughput rates could be increased by a factor of 20 on a single compute node.

\begin{figure}[!htbp]
\centering
\includegraphics[width=0.8\columnwidth]{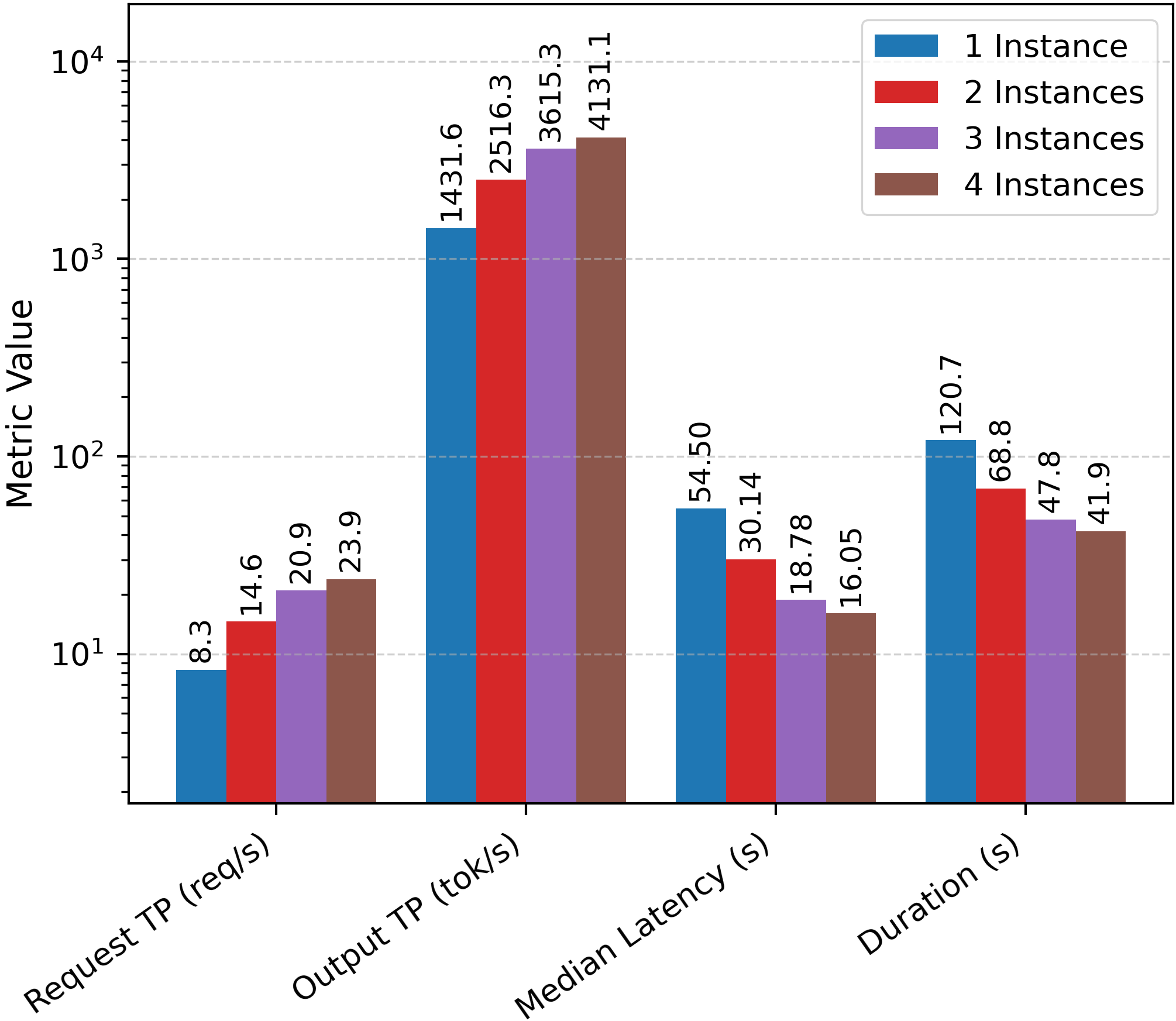}
  \vspace{-0.1in}
\caption{Auto-scaling performance: Single vs.\ two, three, and four instances for Llama 3.3 70B on Sophia's A100 GPUs under maximum load.}
\label{fig:scaling_comparison_70b}
\vspace{-0.15in}
\end{figure}

\textbf{Summary (FIRST vs.\ vLLM Direct):}
FIRST introduces minimal overhead, outperforming direct vLLM access in peak tests. Interpreting this alongside the rate comparison (\Cref{fig:rate_comparison_70b}), FIRST's gateway architecture is better suited for handling sustained high request volumes typical of scientific workloads, offering superior scalability and robustness compared to direct backend access.


FIRST also offers a batch mode optimized for high-throughput, non-interactive tasks. In this mode, requests are processed using vLLM's offline batch mode without serving within a dedicated job allocation. For instance, a batch job processing 1000 requests using the Llama 3.3 70B model achieved an overall output throughput of 2117 tokens/second, completing in 409 seconds. Because batch jobs typically involve a cold start (loading the targeted model), the initial model loading time can dominate the total execution time for smaller batches. However, for larger workloads (>10,000 requests), the amortization of the loading cost across many requests makes batch mode highly efficient for large-scale data generation.



\subsubsection{\textbf{Auto-Scaling Performance (Llama 3.3 70B Model)}}
Auto-scaling in FIRST refers to the capability, configured via Globus Compute endpoint settings (\autoref{sec:gc_endpoint_config}), to automatically launch additional instances (jobs) of a model on separate compute resources when the existing instance(s) become saturated. This capability allows the system to adapt dynamically to increased request loads.
To evaluate FIRST's ability to scale out, we benchmarked the Llama 3.3 70B model using FIRST first on a single instance and then on configurations that automatically launched additional instances (up to four total) when demand saturated existing ones. All scenarios used an infinite request rate with the ShareGPT dataset. 

The results, presented in \Cref{fig:scaling_comparison_70b}, show that scaling from one to multiple instances significantly increased the overall processing capacity. Request throughput scaled from 8.3 req/s (1 instance) to 14.6 req/s (2 instances), 20.9 req/s (3 instances), and 23.9 req/s (4 instances). Similarly, output token throughput increased substantially from 1432 (1 instance) to 4131 tok/s (4 instances). In terms of the output token throughput, we observe an increase to 1.75$\times$ as we scale to two instance, to 2.52$\times$ at three instances, and 2.88$\times$ at four instances. Our overall scaling is currently limited by the ability of Globus Compute to scale and route requests to the multiple instances. Overall, the scaling demonstrates FIRST's capability to effectively leverage additional resources provisioned via Globus Compute's auto-scaling features by distributing the incoming request load across simultaneously running model instances.

\begin{figure}[!htbp]
  \centering
  \includegraphics[width=0.8\columnwidth]{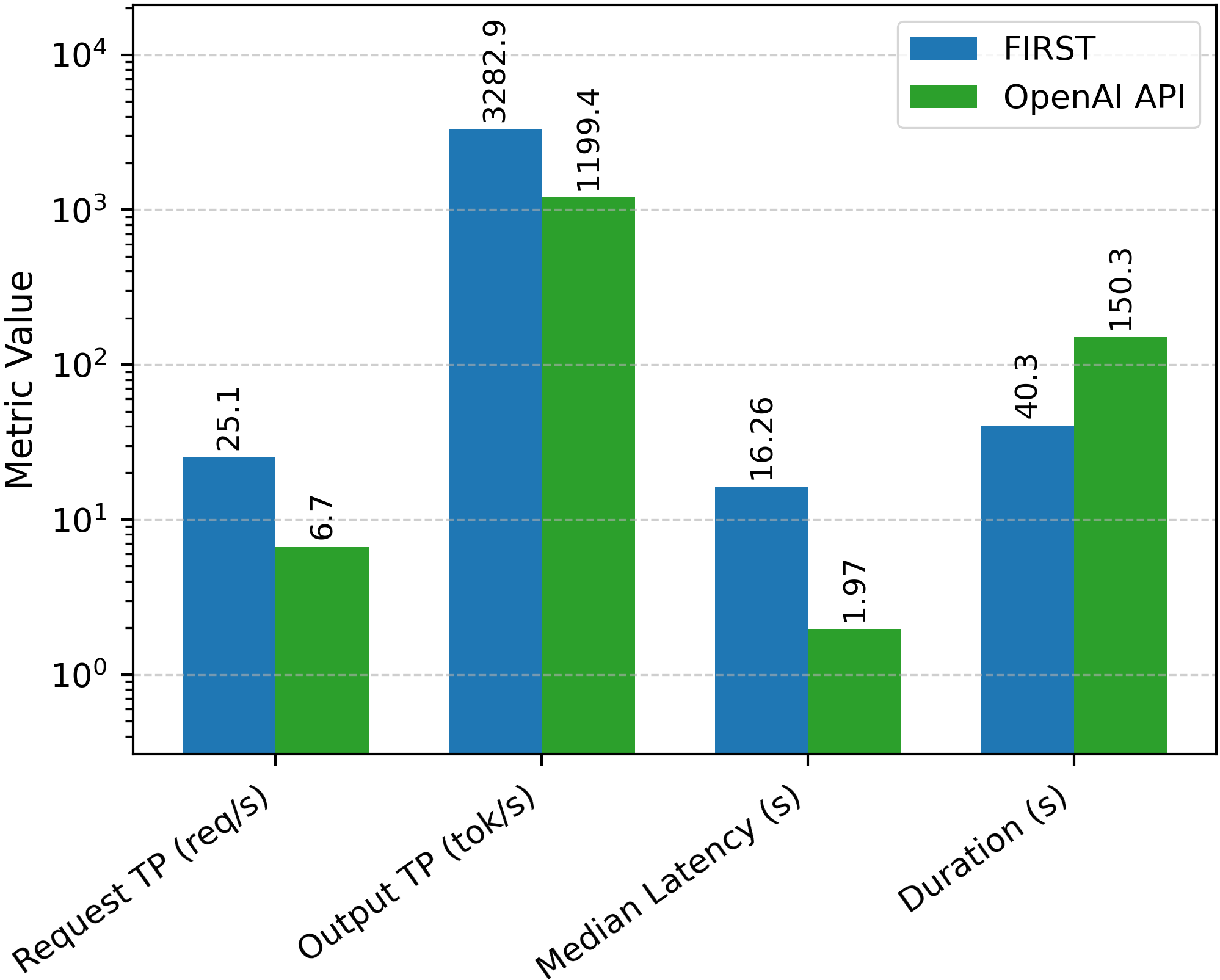}
  \vspace{-0.1in}
  \caption{Benchmark comparison of FIRST (Llama 3.1 8B) vs.\ OpenAI API (the $\sim$8B-parameter GPT-4o-mini). Note: OpenAI API results may be influenced by service-side rate limiting.}
  \label{fig:openai_comparison}
  \vspace{-0.15in}
\end{figure}

Scaling also substantially decreased the median end-to-end latency from 54.5s (1 instance) down to 30.1s (2 instances), 18.8s (3 instances), and finally 16.0s (4 instances) by distributing the load and reducing queueing within each instance. These results validate that the FIRST architecture, combined with Globus Compute's resource management, can effectively auto-scale to handle increased inference demand, improving both throughput and responsiveness.

\subsubsection{\textbf{FIRST vs. OpenAI API}}
To provide context against industry-standard cloud offerings, we compared FIRST (running the open-source Llama 3.1 8B model) with the OpenAI API serving GPT-4o-mini (estimated to be in a similar parameter class). This comparison serves to contextualize FIRST's performance within the broader inference landscape, not to position it as a direct competitor to commercial services. The results are shown in \Cref{fig:openai_comparison}. In this comparison, FIRST significantly outperforms the OpenAI API in request throughput (25.1 vs.\ 6.7 req/s) and output token throughput (3283 vs.\ 1199 tok/s). Conversely, the OpenAI API offers lower median latency (2.0s vs.\ 16.3s for FIRST). This highlights a common trade-off: while commercial cloud APIs such as OpenAI may provide low latency, likely due to significant scale and optimization beyond our visibility, FIRST deployed on dedicated HPC resources demonstrates the capacity to handle a much larger volume of concurrent requests, achieving substantially higher overall throughput. This makes FIRST particularly suitable for scientific workloads involving large numbers of concurrent users or high-volume batch processing tasks, where maximizing throughput on available, secure HPC resources is paramount. The results underscore the value of specialized, self-hosted inference solutions within the HPC ecosystem. It is important to note that benchmarks against the OpenAI API are subject to service-side rate limits and potentially variable resource allocation, which may influence the observed performance.

\subsubsection{\textbf{Concurrency and Throughput Analysis for WebUI}}

We measured the performance of our WebUI deployment under increasing concurrency, defined as the number of simultaneous inference requests being processed. Measurements include request throughput (the number of successful requests completed per second) and output token throughput (the total number of output tokens generated per second across all successful requests). Benchmarks were performed using simulated concurrent WebUI sessions targeting three models: Llama-3.1-8B, Gemma-27B, and Llama-3.3-70B. As shown in Table~\ref{tab:concurrency}, both token and request throughput scale nearly linearly from 50 to 500 concurrent sessions, with diminishing returns beyond this point as the backend approaches saturation. Shorter runs (60 sec) consistently yielded higher throughput than longer runs (120 sec), likely due to reduced resource contention and the avoidance of long-tail latency effects. Overall, the results demonstrate that our web interface can sustain high concurrency without becoming a performance bottleneck.

\section{Case Studies}
We present case studies in which researchers leverage FIRST capabilities to support diverse scientific applications.

\subsection{Model Evaluation and Comparison}

Using the Inference Gateway, researchers benchmarked fifteen GPT‑style models—including AuroraGPT \cite{auroragpt_website} fine‑tuned on HPC documentation and a suite of open‑source LLMs—against custom models trained on proprietary scientific corpora. Over 50,000 inference requests generated more than one million tokens, allowing the team to assess accuracy on domain‑specific terminology, compare architectures of varying size, and quantify the trade‑offs between computational cost and performance. The gateway’s ability to swap models instantly eliminated manual deployment steps, yielding a 40 percent reduction in total evaluation time while preserving consistent throughput across all model variants.

\subsection{HPC Assistant with RAG}
To provide on‑demand support for a high‑performance computing environment, a chatbot was constructed that combines FIRST’s embedding and inference services. NVIDIA’s NV‑Embed‑v2 produced dense vector representations of HPC manuals, guides, and troubleshooting documents, which were stored in a FAISS index \cite{douze2025faisslibrary} for rapid similarity search. When a user poses a question, a Retrieval‑Augmented Generation (RAG) pipeline \cite{10.5555/3495724.3496517} retrieves the most relevant passages and incorporates them into the prompt sent to the LLM. This architecture enables the system to disambiguate user intent, surface highly relevant guidance, and dramatically improve the quality of HPC‑specific assistance.

\subsection{Synthetic Data Generation for Training}
Researchers leveraged the Inference Gateway to generate synthetic data for training machine learning models, showcasing the system's capabilities in supporting large-scale data generation workflows. The batch mode was particularly valuable for this use case, allowing researchers to generate billions of tokens of synthetic data with minimal overhead. The system processed a batch job of 10,000 requests, generating over 1 million tokens of synthetic data. The batch mode eliminated the per-request overhead, achieving a throughput of 25,000 tokens per second per model. The total job completed in under 24 hours, compared to an estimated 2 days using a manual deployment approach. This significant time savings enabled the researchers to iterate more quickly on their data generation strategies and produce larger volumes of training data.

\begin{table}[t]
\centering
\caption{WebUI benchmark results per model. 
Conc = concurrency; TP = throughput tokens; Req = requests.}
  \vspace{-0.1in}
\label{tab:concurrency}
\footnotesize
\setlength{\tabcolsep}{4pt}
\renewcommand{\arraystretch}{1.1}
\resizebox{\columnwidth}{!}{%
\begin{tabular}{lcccccc}
\toprule
\multirow{2}{*}{Model} & \multirow{2}{*}{Conc.} & \multicolumn{2}{c}{60 s} & \multicolumn{2}{c}{120 s} \\
\cmidrule(lr){3-4} \cmidrule(lr){5-6}
 &  & \multicolumn{1}{c}{TP/s} & \multicolumn{1}{c}{Req./s} & \multicolumn{1}{c}{TP/s} & \multicolumn{1}{c}{Req./s} \\
\midrule
\multirow{5}{*}{Llama-3.1-8B} 
 & 50  & 690.68 & 4.97 & 441.17 & 3.12 \\
 & 100 & 738.33 & 5.25 & 563.18 & 4.01 \\
 & 300 & 1103.70 & 7.90 & 981.45 & 6.81 \\
 & 500 & 1672.15 & 12.08 & 1271.04 & 8.94 \\
 & 700 & 2119.50 & 14.68 & 1385.93 & 9.74 \\
\midrule
\multirow{5}{*}{Gemma-27B} 
 & 50  & 297.97 & 2.70 & 864.83 & 5.13 \\
 & 100 & 906.62 & 5.42 & 865.05 & 5.10 \\
 & 300 & 1469.53 & 8.67 & 1211.75 & 7.25 \\
 & 500 & 1849.67 & 10.95 & 1144.79 & 6.83 \\
 & 700 & 2651.40 & 15.57 & 1353.15 & 8.17 \\
\midrule
\multirow{5}{*}{Llama-3.3-70B} 
 & 50  & 217.38 & 1.63 & 472.05 & 3.57 \\
 & 100 & 785.83 & 5.88 & 503.52 & 3.86 \\
 & 300 & 1061.93 & 7.92 & 948.13 & 7.13 \\
 & 500 & 1646.53 & 12.30 & 1176.39 & 8.75 \\
 & 700 & 2134.10 & 15.67 & 1372.27 & 10.35 \\
\bottomrule
\end{tabular}%
}
\end{table}

\balance

\section{Conclusions and Future Work}

We introduced FIRST, an Inference-as-a-Service framework that securely hosts and scales AI inference workloads on HPC systems using Globus for authentication, communication, and scheduler integration. FIRST is flexible, provides both API and web-chat access to AI models, including LLMs and domain specific models, supports multiple inference frameworks, optimizes performance and latency, and offers greater data control compared to commercial cloud solutions. Our work represents a significant step toward democratizing AI access, which enables researchers without deep HPC expertise to leverage advanced inference capabilities while preserving institutional security and resource policies.

In future work we aim to expand FIRST's capabilities to enable direct job submission for users, allowing AI Models to execute custom codes as tool calls and run traditional HPC simulations through the same API interface. We will reduce barriers for new users by providing a unified interface for both AI inference and traditional HPC workloads, further demonstrating how FIRST can coexist with and enhance, rather than replace, existing HPC infrastructure.
We plan to expand our backend API to support scientific foundation models beyond LLMs, including vision models, protein folding models, and other domain-specific AI models commonly used in scientific research. We also plan to improve scheduling for resource optimization, enhance monitoring for deeper insights, and strengthen security to meet evolving threats and compliance requirements.

\begin{acks}
This research used resources of the Argonne Leadership Computing Facility, a U.S. Department of Energy (DOE) Office of Science user facility at Argonne National Laboratory and is based on research supported by the U.S. DOE Office of Science-Advanced Scientific Computing Research Program, under Contract No. DE-AC02-06CH11357.

\end{acks}

\bibliographystyle{ACM-Reference-Format}
\bibliography{sc2025-references,references}


\begin{thebibliography}{39}


\ifx \showCODEN    \undefined \def \showCODEN     #1{\unskip}     \fi
\ifx \showISBNx    \undefined \def \showISBNx     #1{\unskip}     \fi
\ifx \showISBNxiii \undefined \def \showISBNxiii  #1{\unskip}     \fi
\ifx \showISSN     \undefined \def \showISSN      #1{\unskip}     \fi
\ifx \showLCCN     \undefined \def \showLCCN      #1{\unskip}     \fi
\ifx \shownote     \undefined \def \shownote      #1{#1}          \fi
\ifx \showarticletitle \undefined \def \showarticletitle #1{#1}   \fi
\ifx \showURL      \undefined \def \showURL       {\relax}        \fi
\providecommand\bibfield[2]{#2}
\providecommand\bibinfo[2]{#2}
\providecommand\natexlab[1]{#1}
\providecommand\showeprint[2][]{arXiv:#2}

\bibitem[sop({[n.\,d.]})]%
        {sophia}
 \bibinfo{year}{[n.\,d.]}\natexlab{}.
\newblock \bibinfo{booktitle}{\emph{ALCF Sophia}}.
\newblock
\urldef\tempurl%
\url{https://www.alcf.anl.gov/sophia}
\showURL{%
\tempurl}


\bibitem[gar(2025)]%
        {gardens_ai}
 \bibinfo{year}{2025}\natexlab{}.
\newblock \bibinfo{title}{The Gardens.ai Platform}.
\newblock \bibinfo{howpublished}{\url{https://thegardens.ai/}}.
\newblock
\newblock
\shownote{Accessed: 2025-01-10}.


\bibitem[Altair({[n.\,d.]})]%
        {pbs_pro}
\bibfield{author}{\bibinfo{person}{Altair}.} \bibinfo{year}{[n.\,d.]}\natexlab{}.
\newblock \bibinfo{title}{PBS Professional Scheduler}.
\newblock
\urldef\tempurl%
\url{https://altair.com/pbs-professional/}
\showURL{%
\tempurl}


\bibitem[{Amazon Web Services}(2024)]%
        {aws_batch_ml}
\bibfield{author}{\bibinfo{person}{{Amazon Web Services}}.} \bibinfo{year}{2024}\natexlab{}.
\newblock \bibinfo{title}{{AWS Batch for Machine Learning}}.
\newblock \bibinfo{howpublished}{\url{https://aws.amazon.com/batch/}}.
\newblock


\bibitem[{Argonne National Laboratory}({[n.\,d.]})]%
        {auroragpt_website}
\bibfield{author}{\bibinfo{person}{{Argonne National Laboratory}}.} \bibinfo{year}{[n.\,d.]}\natexlab{}.
\newblock \bibinfo{title}{{AuroraGPT}}.
\newblock \bibinfo{howpublished}{\url{https://auroragpt.anl.gov/}}.
\newblock
\newblock
\shownote{Accessed: April 10, 2025}.


\bibitem[{Artillery Project}(2025)]%
        {artillery}
\bibfield{author}{\bibinfo{person}{{Artillery Project}}.} \bibinfo{year}{2025}\natexlab{}.
\newblock \bibinfo{title}{Artillery - Modern load testing and smoke testing for SRE and DevOps}.
\newblock \bibinfo{howpublished}{\url{https://www.artillery.io}}.
\newblock
\newblock
\shownote{Accessed: 2025-04-14}.


\bibitem[Bryant et~al\mbox{.}(2025)]%
        {vllm_issue_12705}
\bibfield{author}{\bibinfo{person}{Russell Bryant} {et~al\mbox{.}}} \bibinfo{year}{2025}\natexlab{}.
\newblock \bibinfo{title}{[RFC]: Scale the API server across multiple CPUs \#12705}.
\newblock \bibinfo{howpublished}{\url{https://github.com/vllm-project/vllm/issues/12705}}.
\newblock
\newblock
\shownote{GitHub Issue}.


\bibitem[Chard et~al\mbox{.}(2020)]%
        {10.1145/3369583.3392683}
\bibfield{author}{\bibinfo{person}{Ryan Chard}, \bibinfo{person}{Yadu Babuji}, \bibinfo{person}{Zhuozhao Li}, \bibinfo{person}{Tyler Skluzacek}, \bibinfo{person}{Anna Woodard}, \bibinfo{person}{Ben Blaiszik}, \bibinfo{person}{Ian Foster}, {and} \bibinfo{person}{Kyle Chard}.} \bibinfo{year}{2020}\natexlab{}.
\newblock \showarticletitle{funcX: A Federated Function Serving Fabric for Science}. In \bibinfo{booktitle}{\emph{Proceedings of the 29th International Symposium on High-Performance Parallel and Distributed Computing}} (Stockholm, Sweden) \emph{(\bibinfo{series}{HPDC '20})}. \bibinfo{publisher}{Association for Computing Machinery}, \bibinfo{address}{New York, NY, USA}, \bibinfo{pages}{65--76}.
\newblock
\showISBNx{9781450370523}
\href{https://doi.org/10.1145/3369583.3392683}{doi:\nolinkurl{10.1145/3369583.3392683}}


\bibitem[Chesneau et~al\mbox{.}({[n.\,d.]})]%
        {gunicorn}
\bibfield{author}{\bibinfo{person}{Benoît Chesneau}, \bibinfo{person}{Paul~J. Davis}, {et~al\mbox{.}}} \bibinfo{year}{[n.\,d.]}\natexlab{}.
\newblock \bibinfo{title}{Gunicorn: Python WSGI HTTP Server for UNIX}.
\newblock \bibinfo{howpublished}{\url{https://gunicorn.org}}.
\newblock


\bibitem[Chitty-Venkata et~al\mbox{.}(2024)]%
        {10.1109/SCW63240.2024.00178}
\bibfield{author}{\bibinfo{person}{Krishna~Teja Chitty-Venkata}, \bibinfo{person}{Siddhisanket Raskar}, \bibinfo{person}{Bharat Kale}, \bibinfo{person}{Farah Ferdaus}, \bibinfo{person}{Aditya Tanikanti}, \bibinfo{person}{Ken Raffenetti}, \bibinfo{person}{Valerie Taylor}, \bibinfo{person}{Murali Emani}, {and} \bibinfo{person}{Venkatram Vishwanath}.} \bibinfo{year}{2024}\natexlab{}.
\newblock \showarticletitle{LLM-Inference-Bench: Inference Benchmarking of Large Language Models on AI Accelerators}. In \bibinfo{booktitle}{\emph{Proceedings of the SC '24 Workshops of the International Conference on High Performance Computing, Network, Storage, and Analysis}} (Atlanta, GA, USA) \emph{(\bibinfo{series}{SC-W '24})}. \bibinfo{publisher}{IEEE Press}, \bibinfo{pages}{1362–1379}.
\newblock
\showISBNx{9798350355543}
\href{https://doi.org/10.1109/SCW63240.2024.00178}{doi:\nolinkurl{10.1109/SCW63240.2024.00178}}


\bibitem[Christie et~al\mbox{.}({[n.\,d.]})]%
        {uvicorn}
\bibfield{author}{\bibinfo{person}{Tom Christie} {et~al\mbox{.}}} \bibinfo{year}{[n.\,d.]}\natexlab{}.
\newblock \bibinfo{title}{Uvicorn: ASGI web server implementation for Python}.
\newblock \bibinfo{howpublished}{\url{https://www.uvicorn.org/}}.
\newblock


\bibitem[Christie and contributors(2025)]%
        {django-rest-framework}
\bibfield{author}{\bibinfo{person}{Tom Christie} {and} \bibinfo{person}{contributors}.} \bibinfo{year}{2025}\natexlab{}.
\newblock \bibinfo{title}{Django REST framework}.
\newblock \bibinfo{howpublished}{\url{https://www.django-rest-framework.org/}}.
\newblock
\newblock
\shownote{Version 3.14.0}.


\bibitem[Contributors(2023)]%
        {sharegpt2023}
\bibfield{author}{\bibinfo{person}{ShareGPT Contributors}.} \bibinfo{year}{2023}\natexlab{}.
\newblock \bibinfo{title}{ShareGPT: Crowdsourced Conversations for Instruction Tuning}.
\newblock \bibinfo{howpublished}{\url{https://huggingface.co/datasets/anon8231489123/ShareGPT_Vicuna_unfiltered}}.
\newblock
\newblock
\shownote{Accessed: 2025-04-11}.


\bibitem[Davis(2021)]%
        {django_ninja_website}
\bibfield{author}{\bibinfo{person}{Joshua M.~L. Davis}.} \bibinfo{year}{2021}\natexlab{}.
\newblock \bibinfo{title}{Django Ninja}.
\newblock
\urldef\tempurl%
\url{https://django-ninja.dev/}
\showURL{%
\tempurl}


\bibitem[Dettmers et~al\mbox{.}(2022)]%
        {dettmers2022llm}
\bibfield{author}{\bibinfo{person}{Tim Dettmers} {et~al\mbox{.}}} \bibinfo{year}{2022}\natexlab{}.
\newblock \showarticletitle{{LLM.int8()}: 8-bit Matrix Multiplication for Transformers at Scale}.
\newblock \bibinfo{journal}{\emph{arXiv preprint arXiv:2208.07339}} (\bibinfo{year}{2022}).
\newblock


\bibitem[Douze et~al\mbox{.}(2025)]%
        {douze2025faisslibrary}
\bibfield{author}{\bibinfo{person}{Matthijs Douze}, \bibinfo{person}{Alexandr Guzhva}, \bibinfo{person}{Chengqi Deng}, \bibinfo{person}{Jeff Johnson}, \bibinfo{person}{Gergely Szilvasy}, \bibinfo{person}{Pierre-Emmanuel Mazaré}, \bibinfo{person}{Maria Lomeli}, \bibinfo{person}{Lucas Hosseini}, {and} \bibinfo{person}{Hervé Jégou}.} \bibinfo{year}{2025}\natexlab{}.
\newblock \bibinfo{title}{The Faiss library}.
\newblock
\showeprint[arxiv]{2401.08281}~[cs.LG]
\urldef\tempurl%
\url{https://arxiv.org/abs/2401.08281}
\showURL{%
\tempurl}


\bibitem[Facility({[n.\,d.]})]%
        {alcf_sophia_docs}
\bibfield{author}{\bibinfo{person}{Argonne Leadership~Computing Facility}.} \bibinfo{year}{[n.\,d.]}\natexlab{}.
\newblock \bibinfo{title}{Sophia Documentation}.
\newblock
\urldef\tempurl%
\url{https://www.alcf.anl.gov/sophia}
\showURL{%
\tempurl}


\bibitem[Facility(2024)]%
        {alcf_polaris_docs}
\bibfield{author}{\bibinfo{person}{Argonne Leadership~Computing Facility}.} \bibinfo{year}{2024}\natexlab{}.
\newblock \bibinfo{title}{Polaris System Documentation}.
\newblock
\urldef\tempurl%
\url{https://www.alcf.anl.gov/polaris}
\showURL{%
\tempurl}


\bibitem[Feil(2023)]%
        {infinity}
\bibfield{author}{\bibinfo{person}{Michael Feil}.} \bibinfo{year}{2023}\natexlab{}.
\newblock \bibinfo{booktitle}{\emph{Infinity - To Embeddings and Beyond}}.
\newblock
\href{https://doi.org/10.5281/zenodo.11630143}{doi:\nolinkurl{10.5281/zenodo.11630143}}


\bibitem[for AI-Innovation({[n.\,d.]})]%
        {llm-serving}
\bibfield{author}{\bibinfo{person}{Center for AI-Innovation}.} \bibinfo{year}{[n.\,d.]}\natexlab{}.
\newblock \bibinfo{title}{llm-serving}.
\newblock
\urldef\tempurl%
\url{https://github.com/Center-for-AI-Innovation/llm-serving}
\showURL{%
\tempurl}


\bibitem[Foster(2011)]%
        {5755602}
\bibfield{author}{\bibinfo{person}{Ian Foster}.} \bibinfo{year}{2011}\natexlab{}.
\newblock \showarticletitle{Globus Online: Accelerating and Democratizing Science through Cloud-Based Services}.
\newblock \bibinfo{journal}{\emph{IEEE Internet Computing}} \bibinfo{volume}{15}, \bibinfo{number}{3} (\bibinfo{year}{2011}), \bibinfo{pages}{70--73}.
\newblock
\href{https://doi.org/10.1109/MIC.2011.64}{doi:\nolinkurl{10.1109/MIC.2011.64}}


\bibitem[{Google Cloud}(2024)]%
        {google_tpu}
\bibfield{author}{\bibinfo{person}{{Google Cloud}}.} \bibinfo{year}{2024}\natexlab{}.
\newblock \bibinfo{title}{{Cloud TPU}}.
\newblock \bibinfo{howpublished}{\url{https://cloud.google.com/tpu}}.
\newblock


\bibitem[{Hugging Face}(2024)]%
        {hf_tgi}
\bibfield{author}{\bibinfo{person}{{Hugging Face}}.} \bibinfo{year}{2024}\natexlab{}.
\newblock \bibinfo{title}{{Text Generation Inference}}.
\newblock \bibinfo{howpublished}{\url{https://github.com/huggingface/text-generation-inference}}.
\newblock


\bibitem[IBM({[n.\,d.]})]%
        {lsf}
\bibfield{author}{\bibinfo{person}{IBM}.} \bibinfo{year}{[n.\,d.]}\natexlab{}.
\newblock \bibinfo{title}{LSF Scheduler}.
\newblock
\urldef\tempurl%
\url{https://www.ibm.com/docs/en/spectrum-lsf/10.1.0/}
\showURL{%
\tempurl}


\bibitem[Jette and Wickberg(2023)]%
        {slurm}
\bibfield{author}{\bibinfo{person}{Morris~A. Jette} {and} \bibinfo{person}{Tim Wickberg}.} \bibinfo{year}{2023}\natexlab{}.
\newblock \showarticletitle{Architecture of the Slurm Workload Manager}. In \bibinfo{booktitle}{\emph{Job Scheduling Strategies for Parallel Processing}}, \bibfield{editor}{\bibinfo{person}{Dalibor Klus{\'a}{\v{c}}ek}, \bibinfo{person}{Julita Corbal{\'a}n}, {and} \bibinfo{person}{Gonzalo~P. Rodrigo}} (Eds.). \bibinfo{publisher}{Springer Nature Switzerland}, \bibinfo{address}{Cham}, \bibinfo{pages}{3--23}.
\newblock
\showISBNx{978-3-031-43943-8}


\bibitem[Kwon et~al\mbox{.}(2023)]%
        {kwon2023efficient}
\bibfield{author}{\bibinfo{person}{Woosuk Kwon}, \bibinfo{person}{Zhuohan Li}, \bibinfo{person}{Siyuan Zhuang}, \bibinfo{person}{Ying Sheng}, \bibinfo{person}{Lianmin Zheng}, \bibinfo{person}{Cody~Hao Yu}, \bibinfo{person}{Joseph~E. Gonzalez}, \bibinfo{person}{Hao Zhang}, {and} \bibinfo{person}{Ion Stoica}.} \bibinfo{year}{2023}\natexlab{}.
\newblock \showarticletitle{Efficient Memory Management for Large Language Model Serving with {PagedAttention}}. In \bibinfo{booktitle}{\emph{Proceedings of the ACM SIGOPS 29th Symposium on Operating Systems Principles}}.
\newblock


\bibitem[Lewis et~al\mbox{.}(2020)]%
        {10.5555/3495724.3496517}
\bibfield{author}{\bibinfo{person}{Patrick Lewis}, \bibinfo{person}{Ethan Perez}, \bibinfo{person}{Aleksandra Piktus}, \bibinfo{person}{Fabio Petroni}, \bibinfo{person}{Vladimir Karpukhin}, \bibinfo{person}{Naman Goyal}, \bibinfo{person}{Heinrich K\"{u}ttler}, \bibinfo{person}{Mike Lewis}, \bibinfo{person}{Wen-tau Yih}, \bibinfo{person}{Tim Rockt\"{a}schel}, \bibinfo{person}{Sebastian Riedel}, {and} \bibinfo{person}{Douwe Kiela}.} \bibinfo{year}{2020}\natexlab{}.
\newblock \showarticletitle{Retrieval-augmented generation for knowledge-intensive NLP tasks}. In \bibinfo{booktitle}{\emph{Proceedings of the 34th International Conference on Neural Information Processing Systems}} (Vancouver, BC, Canada) \emph{(\bibinfo{series}{NIPS '20})}. \bibinfo{publisher}{Curran Associates Inc.}, \bibinfo{address}{Red Hook, NY, USA}, Article \bibinfo{articleno}{793}, \bibinfo{numpages}{16}~pages.
\newblock
\showISBNx{9781713829546}


\bibitem[{Microsoft}(2024)]%
        {azure_hpc_ai}
\bibfield{author}{\bibinfo{person}{{Microsoft}}.} \bibinfo{year}{2024}\natexlab{}.
\newblock \bibinfo{title}{{Azure HPC + AI}}.
\newblock \bibinfo{howpublished}{\url{https://azure.microsoft.com/en-us/solutions/high-performance-computing/}}.
\newblock


\bibitem[NGINX(2004)]%
        {nginx_inc}
\bibfield{author}{\bibinfo{person}{Inc. NGINX}.} \bibinfo{year}{2004}\natexlab{}.
\newblock \bibinfo{title}{NGINX: High Performance Web Server and Reverse Proxy}.
\newblock \bibinfo{howpublished}{\url{https://nginx.org}}.
\newblock
\newblock
\shownote{Accessed April 10, 2025}.


\bibitem[{NVIDIA}(2024)]%
        {tensorrt_llm}
\bibfield{author}{\bibinfo{person}{{NVIDIA}}.} \bibinfo{year}{2024}\natexlab{}.
\newblock \bibinfo{title}{{TensorRT-LLM}}.
\newblock \bibinfo{howpublished}{\url{https://github.com/NVIDIA/TensorRT-LLM}}.
\newblock


\bibitem[{Open WebUI}(2025)]%
        {openwebui}
\bibfield{author}{\bibinfo{person}{{Open WebUI}}.} \bibinfo{year}{2025}\natexlab{}.
\newblock \bibinfo{title}{Open WebUI}.
\newblock \bibinfo{howpublished}{\url{https://openwebui.com/}}.
\newblock
\newblock
\shownote{Accessed: 2025-08-11}.


\bibitem[Prince et~al\mbox{.}(2024)]%
        {Prince2024}
\bibfield{author}{\bibinfo{person}{Michael~H. Prince}, \bibinfo{person}{Henry Chan}, \bibinfo{person}{Aikaterini Vriza}, \bibinfo{person}{Tao Zhou}, \bibinfo{person}{Varuni~K. Sastry}, \bibinfo{person}{Yanqi Luo}, \bibinfo{person}{Matthew~T. Dearing}, \bibinfo{person}{Ross~J. Harder}, \bibinfo{person}{Rama~K. Vasudevan}, {and} \bibinfo{person}{Mathew~J. Cherukara}.} \bibinfo{year}{2024}\natexlab{}.
\newblock \showarticletitle{Opportunities for retrieval and tool augmented large language models in scientific facilities}.
\newblock \bibinfo{journal}{\emph{npj Computational Materials}} \bibinfo{volume}{10}, \bibinfo{number}{1} (\bibinfo{year}{2024}), \bibinfo{pages}{251}.
\newblock
\href{https://doi.org/10.1038/s41524-024-01423-2}{doi:\nolinkurl{10.1038/s41524-024-01423-2}}


\bibitem[Sagi(2025)]%
        {Sagi2025}
\bibfield{author}{\bibinfo{person}{Sriram Sagi}.} \bibinfo{year}{2025}\natexlab{}.
\newblock \showarticletitle{Optimizing LLM Inference: Metrics that Matter for Real Time Applications}.
\newblock \bibinfo{journal}{\emph{Journal of Artificial Intelligence \& Cloud Computing}} \bibinfo{volume}{4}, \bibinfo{number}{1} (\bibinfo{year}{2025}), \bibinfo{pages}{1--4}.
\newblock
\href{https://doi.org/10.47363/JAICC/2025(4)446}{doi:\nolinkurl{10.47363/JAICC/2025(4)446}}


\bibitem[Tuecke et~al\mbox{.}(2016)]%
        {globus_auth}
\bibfield{author}{\bibinfo{person}{Steven Tuecke}, \bibinfo{person}{Rachana Ananthakrishnan}, \bibinfo{person}{Kyle Chard}, \bibinfo{person}{Mattias Lidman}, \bibinfo{person}{Brendan McCollam}, \bibinfo{person}{Stephen Rosen}, {and} \bibinfo{person}{Ian Foster}.} \bibinfo{year}{2016}\natexlab{}.
\newblock \showarticletitle{Globus Auth: A research identity and access management platform}. In \bibinfo{booktitle}{\emph{2016 IEEE 12th International Conference on e-Science (e-Science)}}. \bibinfo{pages}{203--212}.
\newblock
\href{https://doi.org/10.1109/eScience.2016.7870901}{doi:\nolinkurl{10.1109/eScience.2016.7870901}}


\bibitem[{vLLM Project Contributors}(2025)]%
        {vllm_benchmark_script_github}
\bibfield{author}{\bibinfo{person}{{vLLM Project Contributors}}.} \bibinfo{year}{2025}\natexlab{}.
\newblock \bibinfo{title}{{vLLM} Benchmark Serving Script}.
\newblock \bibinfo{howpublished}{\url{https://github.com/vllm-project/vllm/blob/main/benchmarks/benchmark_serving.py}}.
\newblock
\newblock
\shownote{Accessed: April 10, 2025}.


\bibitem[Wallawitsch(2024)]%
        {runpod_vllm_blog}
\bibfield{author}{\bibinfo{person}{Moritz Wallawitsch}.} \bibinfo{year}{2024}\natexlab{}.
\newblock \bibinfo{title}{Introduction to vLLM and PagedAttention}.
\newblock \bibinfo{howpublished}{\url{https://blog.runpod.io/introduction-to-vllm-and-how-to-run-vllm-on-runpod-serverless/}}.
\newblock
\newblock
\shownote{RunPod Blog}.


\bibitem[Wu et~al\mbox{.}(2023)]%
        {10.1145/3581784.3607067}
\bibfield{author}{\bibinfo{person}{Cheng Wu}, \bibinfo{person}{Vinamra Jindal}, \bibinfo{person}{Rohan Dathathri}, \bibinfo{person}{Gregory Diamos}, \bibinfo{person}{Shengen Xie}, \bibinfo{person}{Jeffrey Dean}, \bibinfo{person}{Yacine Jernite}, \bibinfo{person}{Julian Togelius}, \bibinfo{person}{Haidar Patel}, \bibinfo{person}{Gabriel Ilharco}, \bibinfo{person}{Ken Lebowitz}, {and} \bibinfo{person}{Alice Wang}.} \bibinfo{year}{2023}\natexlab{}.
\newblock \showarticletitle{LLM in HPC: Efficient Large Language Model Training and Serving in HPC Systems}. In \bibinfo{booktitle}{\emph{2023 IEEE/ACM Workshop on Machine Learning in High Performance Computing Environments (MLHPC)}}. \bibinfo{pages}{46--56}.
\newblock
\href{https://doi.org/10.1145/3581784.3607067}{doi:\nolinkurl{10.1145/3581784.3607067}}


\bibitem[Zhang et~al\mbox{.}(2023)]%
        {zhang2023flashattention}
\bibfield{author}{\bibinfo{person}{Tri~Dao Zhang} {et~al\mbox{.}}} \bibinfo{year}{2023}\natexlab{}.
\newblock \showarticletitle{{FlashAttention-2}: Faster Attention with Better Parallelism and Work Partitioning}. In \bibinfo{booktitle}{\emph{Advances in Neural Information Processing Systems}}.
\newblock


\bibitem[Zheng et~al\mbox{.}(2024)]%
        {sglang2024}
\bibfield{author}{\bibinfo{person}{Lianmin Zheng}, \bibinfo{person}{Liangsheng Yin}, \bibinfo{person}{Zhiqiang Xie}, \bibinfo{person}{Chuyue Sun}, \bibinfo{person}{Jeff Huang}, \bibinfo{person}{Cody~Hao Yu}, \bibinfo{person}{Shiyi Cao}, \bibinfo{person}{Christos Kozyrakis}, \bibinfo{person}{Ion Stoica}, \bibinfo{person}{Joseph~E. Gonzalez}, \bibinfo{person}{Clark Barrett}, {and} \bibinfo{person}{Ying Sheng}.} \bibinfo{year}{2024}\natexlab{}.
\newblock \bibinfo{title}{SGLang: Efficient Execution of Structured Language Model Programs}.
\newblock
\showeprint[arxiv]{2312.07104}~[cs.AI]
\urldef\tempurl%
\url{https://arxiv.org/abs/2312.07104}
\showURL{%
\tempurl}


\end{thebibliography}

\end{document}